\documentclass[preprint,prb,aps]{revtex4}
\begin{document}

\title{TOWARDS A NEW APPROACH OF QUANTUM DISSIPATION IN SIMPLE CHEMICAL SYSTEMS}
\author{J.P. Badiali} 
\address{LECIME, ENSCP-Universit\'e Pierre et Marie Curie,CNRS/UMR7575   
4 Place Jussieu, 75230 Paris Cedex 05, France}

\begin{abstract}

We propose a new approach for describing the irreversible behavior observed in small quantum systems. It is based on the fact  
that equilibrium thermodynamics may reveal the existence of an underlying dynamics. This is true in the algebraic 
approach of quantum mechanics via the Tomita-Takesaki theorem. A similar result is obtained if we start from 
the path integral expression of the partition function, then an equation of motion is introduced. It 
corresponds to a continuous diffusion process but retaining some aspects of the uncertainty relations. We discuss the relation 
between this equation and the Schr\"{o}dinger equation.
Although the equation of motion is time-irreversible it can be used to describe exactly all the equilibrium properties
of a thermodynamic system. The equilibrium is defined in such a way that there is an exact compensation
between the energy spent, in average, on the paths and the energy put in the system from outside. From this 
equilibrium condition we may give a meaning to the thermal time in path integral formalism. 
In a second step we use the equation of motion to describe irreversible dynamics. 
We show that the equation of motion is equivalent to a quantum Smoluchovski equation. The relation between our Smoluchovski equation
and similar ones used in the literature is emphasized. From the comparison between 
the relaxation time that the system needs to reach its equilibrium state and the thermal time we show in which 
condition the concept of thermostat can be relevant. A standard bistable model is investigated. 
The chemical rate is calculated as a function of time, it appears to be
a non-monotonic function of time. In very particular conditions the stationary value of chemical rate 
can be identified with the Kramers result. 
From all these results it appears that our equation of motion 
or its Smoluchovski version may be considered as a realistic starting point for describing quantum 
dissipation in small systems.


PACS number:03.65Ca, 05.30-d, 05.70-a, 47.53+n .
\end{abstract}
\maketitle
\vspace{0.5cm}

\section{Introduction}
To describe the irreversible evolution of a system the first fundamental element we have at our disposal 
is the second law of thermodynamics from which the so-called arrow of time is introduced. This law 
asserts the existence of a state function, the entropy, that is a non-decreasing function 
of time for any closed system (\cite{landms}). To derive the second law from statistical 
mechanics we have to explain how a macroscopic system may exhibit an irreversible behavior when
the microscopic description is based on time reversible equations (see for instance 
(\cite{zeh})). Boltzmann tried to solve this problem via its derivation of the $H-theorem$ 
that generated an extensive literature (\cite{zeh}). To explain the irreversible behavior 
of chemical reactions but staying in classical physics Kramers (\cite{kramers}) suggested 
to use the Fokker-Planck equation that results from general aspects of the theory of 
random processes (\cite{sacha}). In this equation position and momentum of particles appear 
on the same footing. In some cases it is possible to eliminate 
the momentum and to get a Smoluchovski equation. Different regime can be 
observed depending on the value of the friction coefficient (\cite{kramers}).\\ 
In modern approaches of chemical kinetics the starting point is a quantum description. 
Then we have to deal with the following question: how to describe quantum dissipation 
if we use the Schr\"{o}dinger equation 
that is time reversible in the Wigner sense. Quantum mechanics 
is based on a hamiltonian formalism which can not describe dissipation 
processes (\cite{caldeira}). The system+reservoir methods give a route for introducing irreversibility 
(see for instance (\cite{weiss})) and they have permitted a large improvement in our understanding of the dynamics 
in dense medium. This kind of approach has been extensively used to describe, for instance, the tunneling of electrons through a 
dissipative barrier a very important problem in simple electronic devices (\cite{weiss}).\\
In the system+reservoir approaches the system under investigation is considered as the small open part of a large system. 
The dissipation arises from the energy transfer from the small system to its large environment. 
In such approaches the dynamics is described via a quantum mechanical 
Langevin equation (see for instance (\cite{grabert})) for the relevant operators of the 
reduced system or via a generalized quantum master equation
for the reduced density matrix. Frequently the master equation is a Fokker-Planck type equation (\cite{caldeira}). 
The system+reservoir approach gives a correct description provided 
the system relaxation times are larger than $\tau = \beta \hbar$ where 
$\beta = 1/ k_{B}T$ and $\hbar$ is the Planck constant (\cite{weiss}). 
In absence of reservoir the system is described by the Schr\"{o}dinger equation. \\
At this point it is important to underline the difference between a reservoir considered in the system+reservoir
methods and the usual meaning of resevoir in standard statistical mechanics. In the last case the reservoir is an infinite system
without any specific property, its main role is to fix some external measurable parameters like temperature or
chemical potential. If the reservoir is used to fix the temperature $T$ it is called thermostat. In system+reservoir 
approaches specific interactions exist between the system and its reservoir and by fixing the properties of these interactions
we may reproduce some well known properties of the small system, for instance the 
existence of a brownian motion in the classical limit (\cite{caldeira}). In order to eliminate the reservoir 
an average over the reservoir variables is performed by using the canonical density matrix assuming that
the reservoir is in contact with a thermostat.  
Hereafter we associate to "reservoir" the meaning it has in the sytem+reservoir methods 
while "thermostat" corresponds to the standard concept in statistical mechanics. \\ 
In what follows we want to investigate the irreversible behavior of small systems 
isolated from any reservoir but eventually in contact with a thermostat. In standard chemical kinetics 
we have in mind the description of a reaction such as 
\begin{equation}
A + BC \to AB + C
\label{reaction}
\end{equation}
involving three chemical species $A, B, C$. Our main goal is 
to find an equation for describing the irreversible behavior of such systems. Using the concept of reaction coordinate 
a traditional analysis of (\ref{reaction}) leads to consider the properties of a system in presence of a bistable 
potential (\cite{sacha}). We will retain this kind of approach here and we will analyze the case of a particle inserted 
in a box in which there is a bistable potential. Such a system has been investigated in the seminal paper of 
Kramers (\cite{kramers}) in order to mimic a nuclear fission process. More generally molecular double minimum 
processes have been the subject of many experimental and theoretical 
investigations, in the attempt of a better understanding of elementary chemical reactions. \\

This paper is organized as follows. In Section $2$ we focus on systems at the thermodynamic equilibrium and we 
show that for such systems it exists an underlying dynamics. This idea has been first developped in the 
so-called algebraic approach of quantum physics in relation with the Tomita -Takesaki theorem (\cite{tomita}), 
here a similar idea will be extended to the path integral formalism following an idea suggested by 
Feynman (\cite{feynman}); an equation of motion is derived. We analyze the physical content of this equation; 
in particular we discuss the relation between this equation and the one obtained by introducing an imaginary time in the Schr\"{o}dinger equation. The equation of motion we consider 
is associated with a continuous diffusion process but retaining some aspects of the uncertainty relations. From the entropy  
a thermodynamic equilibrium condition is defined and we give a meaning
to the thermal time in path integral formalism. Our main assumption is that this equation of motion is the general equation
governing the evolution of a particle in an external potential. In Section $3$ we show that the equation of motion 
is equivalent to a quantum Smoluchovski equation; its solution is a density of probability, in which a Boltzmann 
distribution is mixed with a quantity associated with the path integral formalism. We may give a meaning to this 
probability and we discuss the possibility of coupling our system with a thermostat. 
In Section $4$ we use the Smoluchovski equation to describe 
the irreversible evolution of a particle injected in a box in which there is a bistable potential. 
This standard model has been investigated from a general point of view in (\cite{kramers}), (\cite{kampen}) and applied, 
for instance, to describe nuclear fission processes in (\cite{weiden}). For this model we investigate the chemical rate 
and its time dependence. Conclusions are given in Section $5$.  
  
\section{Dynamics in the thermodynamic equilibrium}
Frequently, when a new field of investigation is developing  
different equivalent approaches are proposed. By focusing
on different starting points they may reveal complementary aspects of the theory.
This can be illustrated in the the case of the density matrix. 
As summarized below from the recent developpments of the matrix mechanics elaborated 
by Heisenberg it has been concluded that a given dynamics is associated with the existence 
of the density matrix. We show that an underlying dynamics
can be also associated with the path integral formalism, it is described by an equation of motion. A large
part of this Section will be devoted to an analysis of the physical content of this equation. \\ 
 
\subsection{Thermodynamic equilibrium in algebraic quantum mechanics} 
The matrix mechanics has given rise to the so-called algebraic approach of 
quantum mechanics (\cite{emch}) in which the main point is the existence of 
a C* algebra of non-commutative operators. An important result of this 
approach is the Tomita-Takesaki theorem (\cite{tomita}) (for a recent review 
see (\cite{summers})) that establishes the existence of a 
one parameter group of automorphisms that leaves the algebra  
globally invariant. A consequence of the Tomita-Takesaki theorem is the existence of a 
relation 
\begin{equation} 
(B,(\alpha_{t+i}A)) = ((\alpha_{t}A), B) 
\label{tomita} 
\end{equation}
in which $(,)$ means the inner product and $\alpha_{t}$ is the so-called modular 
group on the algebra. At this level it is impossible to claim that the 
parameter $t$ in (\ref{tomita}) is a time. However we may compare (\ref{tomita}) with the KMS condition 
(\cite{kms}) established by Kubo (\cite{kubo}) and Martin and Schwinger 
(\cite{martin}). Haag and his coworkers (\cite{HHW}) postulated that the KMS 
condition is the correct definition of thermal equilibrium for infinite 
dimensional quantum systems. The KMS condition shows that the correlation 
function between two variables A and B noted $<(\gamma_{t}A)B>$ is analytic 
in the strip $0<Im(t)<\beta\hbar$ and we have 
\begin{equation}
<(\gamma_{t}A)B> = <B (\gamma_{t+i\beta \hbar}A)> 
\label{kms}
\end{equation}
where $\gamma_{t}$ introduces the time translation group defined according to
\begin{equation}
\gamma_{t} A = \exp{\frac{it H}{\hbar}} A \exp{-\frac{it H}{\hbar}}
\label{evolution}
\end{equation} 
in which $H$ is the hamiltonian operator. From the comparison of (\ref{tomita}) and (\ref{kms})
it was concluded that the Tomita-Takesaki theorem is equivalent to the time evolution of the bounded 
operators generated by the Hamilton provided the time is measured in units 
$\beta \hbar$ (\cite{connes}). \\
These results show that the thermal equilibrium characterized by an algebra of operators contains an underlying 
dynamics; these results depend on the form of the density matrix but they do not refer 
explicitly to the existence of the Schr\"{o}dinger equation. 

\subsection{Thermodynamic equilibrium in the path integral formalism}
Starting from the canonical density matrix
\begin{equation}
\rho = \frac{exp-\beta H}{N}
\label{gibbs}
\end{equation}
where $N$ a normalization constant. Feynman (\cite{feynman}) derived a path integral 
formalism for the partition function $Z$. In order to do that it was needed 
i) to start from the Schr\"{o}dinger equation from which we must calculate stationary states and energy 
eigenvalues, ii) to develop some arguments justifying the use of the 
canonical form of the density matrix and iii) to introduce some mathematical 
tricks. Then the partition function can be written as 
\begin{equation} 
Z =\int dx \int Dx(t) \exp{-\frac{1}{\hbar} A_{o}[x(t); 0,\tau ]}
\label{Zfunction}
\end{equation}
in which $Dx(t)$ is the measure for the functional integral and 
$A_{o}[x(t);0,\tau]$ represents the quantity 
\begin{equation}
A_{o}[x(t);0, \tau] = \int\limits_{0}^{\tau} [\frac{1}{2} m 
[\frac{dx(t)}{dt}]^{2} + u(x(t))] dt 
\label{action}
\end{equation} 
calculated on loops $i.e.$ on trajectories on which we have $x(t=0)$ = 
$x(t=\tau)$ = $x$; for each loop the dummy variable, $t$, varies from $0$ to 
$\tau$; $t$ and $\tau$ have the dimension of a time. In (\ref{action}) it is 
assumed that the external potential $u(x(t))$ does not depend explicitly on 
$t$. The integral (\ref{Zfunction}) is calculated on the all values of $x$ 
taken in the volume of the system under consideration. The expression of $Z$ 
given in (\ref{Zfunction}) is just a mathematical transformation of the 
standard expression of $Z=Tr(\exp[-\beta H])$ provided we use $\tau=\beta 
\hbar$. Of course in such a formal derivation we cannot claim that $t$ is a 
time having a physical meaning $i.e.$ that $A_{o}[x(u);0, \tau]$ is an 
action associated with loops in ordinary space time.\\ 
After deriving (\ref{Zfunction}) Feynman (\cite{feynman}) wrote a paragraph entitled 
''Remarks on methods of derivation''on which he suggested that it should be 
possible to derive the expression of $Z$ directly from the description of the motion 
how he did for the quantum mechanical amplitude in quantum mechanics. Why to 
search such a short cut that should avoid the wave function and the energy 
levels ? The Feynman's answer is the following ''in doing that a deeper 
understanding of physical processes might result or possibly more powerful 
methods of statistical mechanics might be evolved''. Hereafter, to be short we will mention these 
remarks as the Feynman's conjecture $(FC)$. From the $FC$ it is suggested 
that $Z=Tr(\exp[-\beta H])$ contains the actual dynamics of the system provided we 
use its path integral version. This is in the same spirit of what has been done in the 
algebraic approach of quantum mechanics. \\
In order to find the dynamics associate with $Z$ we define the quantity \begin{equation} 
q(t_{0},x_{0};t,x) = \smallint 
Dx(t) \exp -\frac{1}{\hbar} A[x(t); t_{0},t] 
\label{q0}
\end{equation} 
in which $A[x(t); t_{0},t]$ is defined in (\ref{action}) but not restricted to close paths. 
From $q(t_{0},x_{0};t,x)$ and a well behaved function $\phi_{0}(x)$ defined 
for $(t = t_{0})$ we may form a real-valued function $\phi(t,x)$ according to 
\begin{equation} 
\phi(t,x) = \smallint \phi_{0}(y) q(t_{0},y;t,x) dy 
\label{fi1} 
\end{equation} 
By using the Feynman-Kac formula, we can see that $\phi(t,x)$ is the 
solution of the equation
\begin{equation} 
-{\partial  \phi(t,x)}/{ \partial t} + {\frac{\hbar}{2m} \Delta_{x} 
\phi(t,x)} -\frac{1}{\hbar}{u(x(t)) \phi(t,x)} = 0 
\label{dif} 
\end{equation}
that verifies the initial condition $\phi(0,x)= \phi_{0}(y)$ $i.e.$ 
$q(t_{0},x_{0};t,x)$ is the fundamental solution of (\ref{dif}) 
in which $\Delta _{x}$ means the laplacian operator taken at the point $x$.\\
Hereafter we consider the equation of motion (\ref{dif}) as the starting 
point of our approach. We cannot proove that (\ref{dif}) is the actual equation of motion but we can 
try to justify it i) by analyzing consequences that we can derive from it and ii) by showing that it is in agreement 
with general trends in modern physics.\\
\subsection{Physical meaning of the equation of motion}
From the canonical form of the density matrix the previous approaches lead to introduce the usual evolution 
operator but with a rescaled time in the case of the algebraic approach or to 
the equation (\ref{dif}) in the case of the path integral formalism. 
In principle these two results do not refer explicitly to the Schr\"{o}dinger 
equation. Indeed, in the spirit of the $FC$ it must be so and (\ref{dif}) must be justified by 
itself from the underlying dynamics. Nevertheless we observe that (\ref{dif}) corresponds formally 
to the Schr\"{o}dinger equation provided we introduce an imaginary-time, this suggests that a link must 
exist between (\ref{dif}) and the Schr\"{o}dinger equation.\\
\subsubsection{Relation with the Schr\"{o}dinger equation}
The imaginary-time approach is based on the observation (\cite{feynman}) that the partition function is formally identical 
to the quantum mechanical amplitude provided the time $t$ is replaced by $it$. By this purely 
formal procedure at least one important question remains 
unsolved: why the solution of (\ref{dif}) that is a real 
valued function has a physical meaning while only the product of the wave function by its 
complex conjugate has a meaning ? We must search a more fundamental relation between (\ref{dif}) and the
Schr\"{o}dinger equation. This is what we have done in (\cite{jpb1}) by focusing on the 
reversible/irreversible behavior of these two equations. To have a self contained paper, hereafter
we shortly recall the physical content of this work; the main point being how to complete the dynamics 
given by (\ref{dif}) in order to describe a reversible behavior. \\
 The Schr\"{o}dinger equation is reversible in the Wigner sense, this means the following.   
If $\psi(t_{1},x))$ is a wave function, its evolution is given by a unitary operator $U(t_{1},t_{0})$  
relating the wave function taken at time $t_{1}$ and the one at time $t_{0}$ according to 
$\psi(t_{1},x) = U(t_{1},t_{0})\psi(t_{0},x)$. If at time $t_{1}$ we consider $\psi^{*}(t_{1},x)$ the complex 
conjugate of $\psi(t_{1},x))$ then by using the same evolution operator and its unitary character we can 
show that for a time interval $(t_{1}- t_{0})$ counted after $t_{1}$ we get a function $\psi^{*}(t_{0},x)$
which is the complex conjugate of $\psi(t_{0},x)$. Thus, to get a Schr\"{o}dinger equation from (\ref{dif}) 
we must introduce a second equation describing a backward motion but with the same fundamental solution as
(\ref{dif}), its solution will be noted $\hat \phi(t,x)$. A fundamental result due to 
Kolmogorov (see (\cite{naga1}) and the references quoted therein) is to show that the equations describing 
forward and backward motions are in duality and from this it is possible to define a Markov 
process for which $\phi(t,x)\hat \phi(t,x)$ is a density of probability. The two real valued 
functions $\phi(t,x)$ and $\hat \phi(t,x)$ can be combined into a complex valued function 
$\psi(t,x)$ verifying a Schr\"{o}dinger equation (\cite{naga1}) and the Born interpretation of 
the wavefunction becomes obvious. Thus we see how from physical arguments we can recover a Schr\"{o}dinger description 
from  (\ref{dif}). \\
Now we have to deal with the following problems: what to say about the dynamics represented by (\ref{dif}) ? 
can we justify this equation ? can we associate to it an acceptable physics ? These problems will be considered in the next subsection but 
detailed arguments already developped in (\cite{jpb1}), (\cite{jpb2}) will be not reported here. 

\subsubsection{The dynamics associated with the equation of motion}
There are many indications showing that spacetime may be discrete rather
continuous (see for instance (\cite{bun}) and the references quoted therein). The choice between 
a discrete and a continuous version of the spacetime structure has already
been analysed by Riemann in the classical world and more than 30 years ago Feynman
presented some doubts concerning the continuum nature of spacetime in the quantum domain
(\cite{fink}). Today it is well accepted that the conventional notions of space and time break down
at the Planck scale where new uncertainty relations have to be introduced. 
A discrete spacetime means that any length is built up from a finite number of the
elementary length, $\delta x$, and any time interval results from a series of individual 'ticks' of
duration $\delta t$. To give a structure to the spacetime we have to introduce a relation between 
$\delta x$ and $\delta t$ this relation determines the scale at which we want to describe the physical world.
At the Planck scale the new uncertainty relations play this role and it is stablished that 
there is a minimum for $\delta x$ and $\delta t$ depending on the velocity of light and
on the gravitation constant. In the
pre-relativistic domain in which we are concerned the only one universal constant that we have is $\hbar$ and for a mass  
$m$ the only one relation that we can introduce between $\delta x$ and $\delta t$ is $\frac{(\delta x)^{2}}{\delta t} = \hbar/m$. 
This relation is a form of the Heisenberg uncertainty relations since we have immediatly $\delta x \delta p = \hbar$ and
$\delta t . \delta E = \hbar /2$ provided we use $\delta p = m (\delta x /\delta t)$ and 
$\delta E = (1/2)m (\frac{\delta x}{\delta t})^{2}$.\\
Although this is probably not needed we assume that the spacetime points
are located on the sites of a regular lattice, as in the chessboard problem investigated in 
(\cite{feynman}), the lattice constants being $\delta x$ and $\delta t$. In absence of external field we assume that the motion 
is as simple as possible. A particle may jump, at random,
from one site to one of its nearest neighbours. This corresponds to a random walk. In absence of extra 
conditions fixing the values of $\delta x$ and $\delta t$ and we may assume that $\delta x$ and $\delta t$ 
tend to zero however in this limit the ratio $(\delta x)^2/(\delta t)$ must remain finite and his value is $\hbar/m$. 
The equation of motion is equivalent to a continuous diffusion process in an external field and the diffusion 
coefficient $D = \hbar/2m$ is related to the existence of uncertainty relations.  

\subsubsection{Properties of the equation of motion}
The solutions of (\ref{dif}) are real valued functions $\phi(t,x)$. 
However since (\ref{dif}) is not a Chapman-Kolmogorov type equation due to the presence 
of the external potential the integration of $\phi(t,x)$ over a finite volume 
is not a time independent quantity and, as a consequence, $\phi(t,x)$ can not be normalized. The function 
$\phi(t,x)$ is a weighted sum of all the paths arriving at the point $x$ and at time $t$ 
when the initial condition $\phi_{0}(y)$ is given. 
The equation (\ref{dif}) is time-irreversible and from it the associated dynamics corresponds to a 
positive semi-group for which we can show (\cite{naga1}) that  
$q(t_{0},x_{0};t,x)$ verifies the law of composition  
\begin{equation} 
q(t_{1},x_{1};t_{2},x_{2}) = \smallint dx_{3} 
q(t_{1},x_{1};t_{3},x_{3}) q(t_{3},x_{3};t_{2}, x_{2}) 
\label{chapkol} 
\end{equation} provided $t_{1} \le t_{3} \le t_{2}$ and therefore it can be used 
to describe the transitions in space-time.\\ 
In (\cite{jpb3}) we have shown that all the averaged quantities calculated 
with (\ref{dif}) on closed loops corresponding to an equilibrium 
situation have a clear physical meaning. 

\subsubsection{Dynamic point of view of the thermodynamic equilibrium}
We have seen that a simple dynamics is associated with (\ref{dif}) . However, in the spirit of the FC, we have to explain how from 
this time irreversible equation we may describe the thermodynamic equilibrium. This is the main goal of this subsection
in which we will define the meaning of the dynamic equilibrium and the physical sense of the thermal 
time in the path-integral formalism. In order to do that we will start from a definition of the entropy since we know 
that the entropy is the corner stone from which we may describe the thermal equilibrium (\cite{callen}) .\\   
Let consider the quantity
\begin{equation} 
S = k_{B} \ln \smallint dx \smallint 
Dx(t) \exp{ - \frac{1}{\hbar} [A_{o}[x(t);0, \tau] - \tau U] }. 
\label{gam} 
\end{equation}
in which $\tau$ is, for the moment, an undetermined free parameter having the dimension of a time and $U$ 
is the external energy put into the system during its preparation, from a thermodynamic point of view $U$ 
corresponds to the internal energy of the system. In (\ref{gam}) the quantity 
$\tau U$ has the meaning of an external action. If it 
exists only one possible trajectory for which the euclidean action $A_{o}[x(t);0, \tau]$ 
exactly compensates the external action $\tau U$ we may say that there is no disorder in 
the system. For real situations it exists a lot of paths for which the order of magnitude of 
$A_{o}[x(t);0, \tau] - \tau U$ is approximately $\hbar$, all these 
trajectories contribute to $S$. Larger is this number of trajectories 
smaller is the order in spacetime and larger is $S$. The calculation 
of $S$ requires to start from a point $x = x(t=0)$ in space and to explore during a 
time interval $\tau$ all the loops around this point and, finally, to perform the same procedure 
for each value of $x$ in the volume of the sample. It is clear 
that $S$ characterizes the order or the disorder in spacetime, but at this level $S$ is not the 
thermal entropy. \\
The quantity $S$ defined above depends on two external parameters 
$\tau$ and $U$. From the definition (\ref{gam}) we may calculate the 
derivative $\frac {dS}{dU}$. In (\cite{jpb2}), it has been shown that we have 
\begin{equation} 
 \frac{\hbar}{k_{B}} \frac{dS}{dU} = 
\tau + [ U - \smallint dx [<u_{K}(x)>_{path}+ <u_{P}(x)>_{path}]\frac{d\tau}{dU} 
\label{to} 
\end{equation}
in which $<u_{K}(x)>_{path}$ is the regular part of the mean value of the 
kinetic energy calculated over the paths localized around the 
initial point $x$ and $<u_{P}(x)>_{path}$ is a similar quantity but 
associated to the potential energy coming from the external potential. From 
(\ref{to}) we may introduce an equilibrium condition
\begin{equation} 
U = \smallint dx [<u_{K}(x)>_{path}+ <u_{P}(x)>_{path}]
\label{equi}
\end{equation}
This condition means that the mean value of the energy calculated on the paths during a time interval $\tau$ 
is equal to the internal energy needed to create the system, $i.e.$ in average, 
we can not spend more energy than the energy put initially in the system. Such an approach 
led to the concept of thermodynamic time in general relativity (\cite{rove1}). 
At the thermal equilibrium we know that $\frac {dS}{dU}$ is a measurable parameter
of the system that corresponds to the reverse of the temperature and from (\ref{to}) and (\ref{equi}) 
we get $\tau = \beta \hbar$. Thus the equilibrium condition (\ref{equi}) lead to introduce a particular 
time scale whic corresponds to the thermal time. 
Note that between $t=0$ and $t = \tau$ no entropy is created in the system as shown in (\cite{jpb4}).
When this value of $\tau$ is introduced into the expression of the entropy we recover exactly all the 
thermodynamic results in the path integral formalism (\cite{jpb3}).\\ 
Clearly $\tau$ is a characteristic of the equilibrium state, it is not surprising that this 
unit of time also gives the scale of time 
in the Tomita-Takesaki theorem that it is concerned with equilibrium situations. 
Thus, although the motion on the paths are described by a time irreversible equation 
we can describe a thermal equilibrium via (\ref{equi}).
Note that $\tau$ does not represent the relaxation time that 
a system taken in a non equilibrium state requires to reach its equilibrium state. 
Such a relaxation time will be investigated
in Section$4$. Another example of relaxation towards an equilibrium state has been given in papers 
devoted to the derivation of a $H-theorem$ (\cite{jpb1}),(\cite{jpb5}). \\ 
Another meaning of $\tau$ can be given from the time-energy uncertainty relation (\cite{jpb4}). If $t < \tau$ the quantum 
fluctuations are larger than $k_{B}T$, $i.e.$ the typical value of the thermal energy and we can not say that there
a well defined thermodynamics for such short times. This gives a quantitative meaning to a very well known idea (\cite{landms}).

\subsubsection{Our main assumption}
In this Section we have seen that (\ref{dif}) can be considered as 
describing a continuous diffusion process retaining some aspects of the uncertainty relations. Although this equation of 
motion is time-irreversible it allows us to reobtain all the equilibrium properties via the equilibrium condition 
(\ref{equi}). In previous papers we have seen that (\ref{dif}) can be completed by a second equation if
we want to describe a reversible process, this establishes a link between (\ref{dif}) and the Schr\"{o}dinger equation based 
on physical arguments.
In (\cite{jpb1}) we have shown that all the quantities calculated at equilibrium on the paths have a physical meaning.
Thus (\ref{dif}) is in agreement with everything we know concerning the thermal equilibrium, 
hereafter we will assume that (\ref{dif}) can be also used to describe non-equilibrium state.  
First, we will show that (\ref{dif}) is also equivalent to a quantum Smoluchoski equation.   

\section{A quantum Smoluchovski equation}
Due to the presence of the potential $u(x)$ in (\ref{dif}),
the solution of this equation is not a density of probability.  
However we may use a transformation introduced by van Kampen (\cite{kampen}) to 
establish a relation between (\ref{dif}) and a Smoluchovski equation and then 
to define a density of probability. 

\subsection{Derivation of the Smoluchovski equation}
Let consider the quantity
\begin{equation} 
P(t,x) = \phi(t,x) \exp{\frac{ - \theta V(x)}{2}}   
\label{pxt}
\end{equation}
in which a potential $V(x)$ is introduced and $(\theta)^{-1}$ is a scale for this potential. \\
Since $V(x)$ is assumed to be independent of $t$ we have 
\begin{equation}
\frac{\partial P(t,x)}{\partial t} = \frac{\partial \phi(t,x)}{\partial t}\exp{\frac{ - \theta V(x)}{2}}    
\label{dpdt}
\end{equation}
By replacing $\frac{\partial \phi(t,x)}{\partial t}$ by its value obtained from (\ref{dif}) and using 
simple mathematical transformations it is possible 
to get the following Smoluchovski equation
\begin{equation}
\frac{\partial P(t,x)}{\partial t} = 
\frac{\hbar}{2m} \Delta_{x}P(t,x) - \frac{\hbar}{2m} \nabla_{x}((\nabla_{x}\theta V(x))P(t,x))
\label{FP}
\end{equation}
in which $V(x)$ is the solution of    
\begin{equation}
\frac{1}{2}\Delta_{x}\theta V(x) - \frac{1}{4}(\nabla_{x} \theta V(x))^{2} + 
\frac{2m}{(\hbar)^{2}} U(x) = 0
\label{defV}
\end{equation}
Finally we introduce the quantity $Z(x,t)=\exp{\frac{ - \theta V(x)}{2}}$, it
verifies the following equation
\begin{equation} 
 {\frac{\hbar}{2m} \Delta_{x} 
Z(t,x)} -\frac{1}{\hbar}{u(x(t)) Z(t,x)} = 0 
\label{eqv} 
\end{equation} 
which is nothing else than (\ref{dif}) in the stationary regime.
Thus from $\phi(t,x)$ we can create a probability $P(t,x)$ given by (\ref{pxt}) which depends on the 
number of paths arriving at $x$ via $\phi(t,x)$ and on the value of the potential $V(x)$ through
$\exp-\frac{\theta V(x)}{2}$. This is a non-traditional expression for a density of probability and we have to 
give a meaning to $P(t,x)$.

\subsection{The meaning of $P(t,x)$}
To solve an equation like (\ref{dif}) 
we separate the variables $x$ and $t$ searching a set of solutions of the form 
$\phi_{n}(t,x) = f_{n}(t) \varphi_{n}(x)$. The time-dependent functions are given by 
$f_{n}(t)= \exp(-\frac{E_{n}- E_{0}}{\hbar}t)$ where $E_{0}$ is the energy of the 
fundamental state and $E_{n}$ is an eigenvalue of the equation 
\begin{equation}
\frac{(\hbar)^{2}}{2m} \Delta_{x} 
\varphi_{n}(x) + ((E_{n}- E_{0}) -u(x))( \varphi_{n}(x) = 0 
\label{phin}
\end{equation}
which is identical to a stationary Schr\"{o}dinger equation for a particle in presence of the initial
external potential $u(x)$. Note that the functions $f_{n}(t)$ are monotonic 
decreasing functions of time in contrast with the solutions of the Schr\"{o}dinger equation that should be   
oscillary functions of time. Using the closure relation between the eigenfunctions we 
can write the fundamental solution of (\ref{FP}) as
\begin{equation}
P(0,y;t,x) = \varphi_{0}(x)\Sigma[\frac{\varphi_{n}(y)}{\varphi_{0}(y)}] \varphi_{n}(x)\exp(-\frac{E_{n}- E_{0}}{\hbar}t)
\label{solfond}
\end{equation} 
where the sum runs on the all values of $n$ from $0$ to infinity and $P(0,y;0,x) = \delta(x-y)$ as a consequence of the closure relation. 
If $f(y)$ is the initial distribution the we have
\begin{equation}
P(t,x) = \smallint P(0,y;t,x) f(y) dy
\end{equation}
If $t$ tends to infinity the limit of $P(t,x)$ is $ P_{eq}(x)= \varphi_{0}(x)^{2} = c \exp -\theta V(x)$. In agreement with 
(\cite{kramers}), (\cite{weiden}) and (\cite{kampen}) we can interpret $P_{eq}(x)$ as the Boltzmann equilibrium density of probability
to be at the point $x$ where there is an effective external potential $V(x)$, $c$ is a normalization constant. 
More generally we can rewrite $P(t,x)$ as
\begin{equation}
P(t,x) = (\varphi_{0}(x))^{2}\Sigma c_{n}(\frac{\phi_{n}(x,t)}{\phi_{0}(x)}) = P_{eq}(x) \gamma(t,x) 
\label{newpxt}
\end{equation}
Thus, before reaching an equilibrium situation, $P(t,x)$ is the product of a Boltzmann distribution 
by a quantity $\gamma(t,x)$ associated with the paths.   
In $\gamma(t,x)$, for each state $n$, we first compare the number of paths arriving at $x$ 
via $\phi_{n}(x,t)$ to the same number at equilibrium given by $\phi_{0}(0,x) = \varphi_{0}(x)$, then we perform
a sum over the states weighted by 
\begin{equation}c_{n}= \smallint \frac{\varphi_{n}(y)}{\varphi_{0}(y)})f(y) dy
\label{cn}
\end{equation} 
If at a given time $t$ there is no path arriving at the point $x$ then $P(t,x)$ vanishes and there is no chance to find a particle
at this time in this point. 
From all the arguments developed above we may interpret $P(t,x)$ as the density of probability to be at the point 
$x$ at the time $t$ taking into account that the initial distribution is given by $f(y)$. \\
If a Smoluchovski or a Fokker-Planck equations have been derived in the system+reservoir methods, 
these equations are based on ingredients including explicitly some properties of the reservoir such as
the frequency spectrum for the reservoir oscillators or the friction coefficient induced by the reservoir particles.
The philosophy in the system+reservoirs methods is to fit the system reservoir 
interactions in order to reproduce, for instance, the brownian motion for the system in the classical limit (\cite{caldeira}). 
In the present work we have no such a reservoir and accordingly no such parameters are present in our Smoluchovski equation. 
In particular there is no temperature in our equation in contrast with what happens in equation based on 
the system+reservoir approach or in the classical approach of Kramers (\cite{kramers}). 
The possibility of introducing the temperature will be analyzed in the next Section.   

\subsection{Contact with a thermostat}
The equations (\ref{FP}) and (\ref{defV}) show that only the quantity $\theta V(x)$ solution of (\ref{eqv})  
is relevant in the calculation of $P(x,t)$.  
In (\cite{kramers}), (\cite{weiden}) the system is assumed to be in contact with 
a thermostat and $\theta = \beta$ has been chosen. To analyse the relevance of the thermostat concept 
in our approach we have to compare the system relaxation time and the thermal time $\tau = \beta \hbar$.
From (\ref{solfond}) and provided that $E_{n}$ is an increasing function of $n$ 
the relaxation time with which $P(t,x)$ will reach its equilibrium value $P_{eq}(x)$ 
is given by $\frac{\hbar}{(E_{1} - E_{0})}$. \\
If $\frac{\hbar}{(E_{1} - E_{0})} << \beta \hbar$ or $k_{B}T <<(E_{1} - E_{0})$
the relaxation of $P(t,x)$ takes place in a regime for which the quantum fluctuations are much larger than the thermal ones;
in such conditions the concept of thermostat is irrelevant. Then we have to consider the system as isolated or possibly submitted 
to a distribution of initial conditions. \\
In the reverse situations $i.e$ if $\frac{\hbar}{(E_{1} - E_{0})} >> \beta \hbar$ 
we may assume that the system in contact with a thermostat and we have to take a thermal average of (\ref{solfond}). This can be realized 
by introducing in (\ref{solfond}) for each value of $n$ an extra thermal factor given by
\begin{equation} 
\frac{\exp(- \beta E_{n})}{\Sigma \exp(- \beta (E_{n})} = \frac{\exp(- \beta (E_{n}- E_{0}))}{\Sigma \exp(- \beta (E_{n}- E_{0}))} = 
\frac{\exp(- \beta (E_{n}- E_{0}))}{Z_{0}} 
\label{thermo}
\end{equation}
in which we have introduced the partition function 
\begin{equation}
Z_{0} = \Sigma \exp(- \beta (E_{n}- E_{0}))
\label{part}
\end{equation}
Due to the thermal effects  $P(0,y;t,x)$ has to 
be replaced by $[P(0,y;t,x)]_{th}$ that we can write as   
 \begin{equation}
[P(0,y;t,x)]_{th} = \frac{1}{Z_{0}}\varphi_{0}(x)\Sigma[\frac{\varphi_{n}(y)}{\varphi_{0}(y)}] 
\varphi_{n}(x)\exp(-\frac{E_{n}- E_{0}}{\hbar}(t + \beta \hbar))
\label{solther}
\end{equation} 
We see from (\ref{solther}) that the thermal effects change $P(0,y;t,x)$ in two ways; first $P(0,y;t,x)$ is rescaled by the partition function
$Z_{0}$ and second the time dependence is shifted by the thermal time. \\
The main result of this Section is to show that the equation of motion (\ref{dif}) is equivalent to a quantum Smoluchovski equation
(\ref{FP}) in an external potential given by (\ref{defV}). The solution of this equation, $P(t,x)$, gives the probability
to be at $x$ for the time $t$. In the next Section we use this equation to investigate an example.  
   
\section{Dynamics in a bistable potential}
In order to illustrate our approach we study the dynamics of a particle injected in a box in which there is a bistable potential.
This is a standard model investigated in general in (\cite{kramers}), (\cite{kampen}) and applied, for instance, to describe the 
nuclear fission (\cite{weiden}). \\
Let be a one dimensional box located in the interval $-b \le x \le b$, in the 
region $ -a \le x \le a$ with $ a < b $ there is a repulsive barrier of height $U_{1}$ while in 
the remaining intervals $[-b,-a]$ and $[a,b]$ it exists an attractive potential of magnitude $U_{0}$. In addition, 
located at $x= \pm b$ we put an infinite repulsive barrier. Thus the potential $u(x)$ introduced in 
(\ref{dif}) is defined by four parameters $(U_{0},U_{1} ;a, b)$. Our first task is to calculate the quantity $\theta V(x)$.

\subsection{Properties of $\theta V(x)$} 
The functions
$\varphi_{n}(x)$ are obtained by the usual methods of quantum mechanics and $\theta V(x)$ can be calculated from 
$\varphi_{0}(x)$. This function is given by $\varphi_{0}(x)= A \cosh K_{0}x$ if $ -a \le x \le a$ and 
$\varphi_{0}(x)= B \sin k_{0}(b - x))$ if $ a \ge x \le b$ the values of the coefficient have been given in (\cite{kampen}). 
It is easy to see that $\theta V(x)$ has a local maximum at $x=0$ given by 
$ \theta V(0) = (2 \pi \alpha \frac{a}{c} + 2\ln \alpha)$ 
with $\alpha^{2} = (\frac{U_{1}}{- U_{0}})$ and $c = (b-a)$. $\theta V(x)$ exhibits two minima located at $\pm {x_{0}}= \pm (a+b)/2$ 
corresponding to $\theta V(x_{0}) =- 2 \ln2$.
The shape of $\theta V(x)$ is reminiscent of the one of $u(x)$ but the ratio
$\frac {V_{0}}{V_{x_{0}}}= (\frac{\pi \alpha \frac{a}{c} + \ln \alpha}{\ln 2})$ is not a simple function of 
$\alpha$ but it also depends on the spatial distribution of $u(x)$ through the parameters $a$ and $c$. 
Near $x=0$ and $x= x_{0}$, $\theta V(x)$ can be expanded according to
\begin{equation}
\theta V(x)= \theta V(0) - K_{0}^{2}x^{2} = \theta V(0) - \theta \frac{1}{2}( 2\pi \Omega_{0})^{2}x^{2}
\label{expV1}
\end{equation}
\begin{equation}
\theta V(x)= \theta V(x_{0}) - k_{0}^{2}(x - x_{0})^{2} = \theta V(x_{0}) - \theta \frac{1}{2}( 2\pi \omega_{0})^{2} (x - x_{0})^{2}
\label{expV2}
\end{equation}
showing the oscillatory character of the potential near the extrema; the relation between $K_{0}$ and $k_{0}$ and the spatial frequency $\Omega_{0}$ 
and $\omega_{0}$ are the same as the ones used by Kramers (\cite{kramers}).

\subsection{The chemical rate}
We can calculate the chemical rate from the definition (\cite{weiden})
\begin{equation}
k(t) = -(\frac{\hbar}{2m}) \frac{1}{P(t)} \frac{dP(t)}{dt}
\label{rate}
\end{equation}
in which 
\begin{equation}
P(t) = \smallint P_{eq}(x) \Sigma c_{n}(\frac{\phi_{n}(x,t)}{\phi_{0}(x)})dx
\label{p1}
\end{equation}
the integration is performed on the domain $ -b \le x \le {0}$. To have a simple model we assume that the initial distribution is
centered at $x=-(b+a)/2$ and its extension, $\sigma$, is very small in comparison with $(b-a)$. \\ 
In the stationary regime $i.e$. for very large
values of $t$ the time-dependent part 
of $P(t)$ is restricted to $P_{1}(t)= \smallint P_{eq}(x) c_{1}(\frac{\phi_{1}(x,t)}{\phi_{0}(x)})dx$
and the rate constant is given by $k_{sta} = \frac{E_{1} - E_{0}}{\hbar}$ which is nothing else than the reverse of the relaxation time 
introduced in $Section 3C$.
By using the values of $(E_{1} - E_{0})$ given in (\cite{kampen}) and the value of $\theta V(0)$ we obtain 
\begin{equation}
k_{sta} = \frac{1}{\pi} \frac{\hbar}{2m} K_{0}. k_{0} \exp{- \theta V(0)} = 2\pi (\frac{\hbar}{2m})\theta \Omega_{0} .\omega_{0} \exp{- \theta V(0)}
\label{ratesta}
\end{equation}
This result looks like the one obtained by Kramers since $k_{sta}$ is determined by the 
product $\Omega_{0} . \omega_{0}$ of the frequecies of the potential 
near extrema (see (\ref{expV1}) and (\ref{expV2}))multiplied by $\exp{- \theta V(0)}$ that we can considered as the 
probability to cross the barrier. Indeed (\ref{ratesta}) becomes identical
to the Kramers result(\cite{kramers}) if we take $\theta = \beta$ and and choose the friction coefficient $\eta$ to be such as 
$\frac{1}{\eta} = \frac{\hbar}{2m}\theta$ of course in this case   
the Smoluchovski equation used here becomes identical to the one investigated by Kramers (equation (11) in (\cite{kramers})). 
Nevertheless, as already noted in (\cite{weiden}), this result 
is not trivial because Kramers used a phenomenological trick (\cite{kramers}), (\cite{sacha}) 
instead of solving the Smoluchovski equation. \\
However it is important to note that (\ref{ratesta}) is not identical to the one of Kramers in general. In the equation of Kramers as well as 
in the Fokker-Planck equations deduce from the system+reservoir approaches, the temperature appears at the level of the Smoluchovski equation. In our approach (\ref{ratesta}) gives the rate constant for an isolated system. The couplig with a thermostat is relevant provided we have $\frac{\hbar}{(E_{1} - E_{0})} >> \beta \hbar$ as discussed in Section $3C$ then we may replace $k_{sta}$ by its thermal average given by 
\begin{equation}
[k_{sta}]_{th} = [\frac{\exp(- \beta (E_{1}- E_{0}))}{Z_{0}}]\frac{E_{1}- E_{0}}{\hbar}
\label{rateth}
\end{equation}  \\
Our result is richer than the one of Kramers since we have an exact solution of the Smoluchovski equation and 
consequently are able to study the time dependence 
of the chemical rate. If we take into account the next terms in the expansion of $P(t)$, it 
is easy to verify that we have $\frac{dP(t)}{dt} \le 0$ in going towards the stationary regime 
showing that we reach $k_{stat}$ by decreasing 
values of $k(t)$ (see Appendix A). In order to to calculate the initial value $k(0)$ we start from (\ref{p1}), we take its time derivative 
and we replace ${\partial  \phi(t,x)}/{ \partial t}$ by its value from (\ref{dif}). After an integration 
by part and taking into account the boundary conditions we get
\begin{equation}
[\frac{dP(t)}{dt}]_{t=0} = \frac{\hbar}{2m} \varphi_{0}(0) [\nabla_{x} \phi(t,x)]_{[x=0; t=o]} = 
\frac{\hbar}{2m} \varphi_{0}(0)[\nabla_{x} f(x)]_{x=0} 
\label{ko}
\end{equation}
The last equality holds from the fact that $\phi(t,x)$ is the solution of (\ref{dif}) verifying the initial condition. 
Since we have assumed that $f(x)$ is entirely localized on the left part of $x=0$, we have $k(0)= 0$. Thus the previous results
show that $k(t)$ is a non monotonic function of $t$. \\

\section{Conclusions}
The main goal of this paper was to find an equation from which we may describe the irreversible 
behavior of small quantum systems. In order to do that we do not try 
to introduce a new kind of quantization as it has been proposed in the past (\cite{weiss}) but we 
start from new and unexpected relations between dynamics and thermodynamic equilibrium. 
From the expression of the partition function we can extract a dynamics by using the algebraic 
approach of quantum mechanics or the path integral formalism. In the first case the dynamics is characterized by 
the usual evolution operator provided a rescaling of time is introduced while in the second approach a time irreversible
partial differential equation is obtained. In Section $2$ we have analyzed the physical content of this equation, in particular
the relation between this equation of motion and the Schr\"{o}dinger equation has been investigated. We have shown that
the equation of motion describes a continuous diffusion process in which we keep in mind some aspects of the uncertainty relations. Although
the equation of motion is time-irreversible it can be used to describe the thermal equilibrium and all the exact
results concerning the thermodynamic quantities are obtained. This is based on an equilibrium definition that asserts that 
the equilibrium is established when the energy spent on the paths is equal to the energy put in the system from outside.
From this equilibrium condition the thermal time $\tau$ is introduced as the time on which we must 
explore the closed paths in order to have the equilibrium. This thermal time is a characteristic 
of the equilibrium. Till this point all the results investigated concern equilibrium properties. Our main assumption
is that the equation of motion can be also used to describe the irreversible dynamics of systems $i.e.$ that the 
equation of motion reveals the general dynamics of real systems. This remains an assumption since initially this 
equation is based on results associated with equilibrium states. We have established that 
the equation of motion is equivalent to a Smoluchovski equation. The density of probability, $P(t,x)$, solution of this 
equation appears as the product of a Boltzmann distribution by a quantity counting the paths and depending on the initial 
system preparation. Nevertheless 
a clear meaning of $P(t,x)$ can be given. For simple systems considered here we may calculated 
the relaxation time $\tau_{r}$ that the system needs to reach an equilibrium when it is initially in a non equilibrium state. We have shown 
that the concept of thermostat is relevant provided we have $\tau_{r} > \tau$. However our results are not restricted 
by such a condition. We have used the Smoluchovski equation to investigate a standard model: the dynamics of a particle injected in a box
in which it exists a bistable potential. We have calculated the chemical rate and we have shown that the chemical rate is 
a non-monotonic function of time for the investigated process. \\

In order to conclude we may present this work on a purely deductive manner, in the spirit of the Feynman conjecture. 
Starting from a primarily discrete space-time we may introduce an equation of motion that retains 
some aspects of the uncertainty relations, this is an easy task since we may use the continuous limit
in our level of physical description. This time-irreversible equation is considered as the actual 
equation of motion in presence of an external field. This equation can be used to describe exactly 
the thermodynamic equilibrium provided we consider a condition of equilibrium. This equation can be also used 
to describe irreversible situations, in this case we may transform it in a quantum Smoluchovski 
equation for which the density of probability solution of this equation has a clear meaning. 
Finally, it is possible to implement this approach by adding to the equation of motion an 
equation describing a backward motion but with the same fundametal solution. By forcing the system in such 
a way we describe a process that is reversible and our equation of motion 
is transformed into a Schr\"{o}dinger equation.     

\section{Appendix}
In Section $4$ we have restricted $P(t)$ to $P_{1}(t)$ in order to get $k_{sta}$ the stationary value of the 
chemical rate. To calculate the first deviation from the stationary state let consider
\begin{equation}
P_{2}(t)= \smallint P_{eq}(x) c_{1}(\frac{\phi_{1}(x,t)}{\phi_{0}(x)})dx + 
\smallint P_{eq}(x) c_{2}(\frac{\phi_{2}(x,t)}{\phi_{0}(x)})dx  
\label{a1}
\end{equation}
We know (\cite{kampen}) that $\phi_{0}(t,x)$ and $\phi_{2}(t,x)$ are symmetric functions of $x$ 
while $\phi_{1}(t,x)$ is antisymmetric. Thus, in the domain $x \in [-b,a]$ the ratio 
$(\frac{\phi_{1}(x,t)}{\phi_{0}(x)})$ is negative and $(\frac{\phi_{2}(x,t)}{\phi_{0}(x)})$ is positive.
The constant $c_{1}$ and $c_{1}$ are determined by the same ratios as shown in (\ref{cn}). 
We can rewrite $P_{2}(t)$ as 
\begin{equation}
P_{2}(t)= a_{1} \exp{- \frac{(E_{1} - E_{0})}{\hbar}t} + a_{2} \exp{- \frac{(E_{2} - E_{0})}{\hbar}t}  
\label{a2}
\end{equation}
in which $a_{1}$ and $a_{2}$ are positive numbers. The derivative of $\frac{-1}{P_{2}(t)} \frac{dP_{2}(t)}{dt}$ can be 
easily obtained and we get 
\begin{equation}
\frac{d}{dt}[\frac{-1}{P_{2}(t)} \frac{dP_{2}(t)}{dt}] = - \frac{1}{P_{2}(t)^{2}}a_{1}a_{2} (E_{2} - E_{1})
\exp{- \frac{(E_{1} - E_{0})}{\hbar}t} \exp{- \frac{(E_{2} - E_{0})}{\hbar}t}
\end{equation}
which is a negative quantity showing that $k(t)$ decreases before reaching its stationary value.


\end{document}